\documentclass[11pt,a4paper]{article}  
\usepackage{jheppub}
\usepackage{bm}
\usepackage{dcolumn}
\usepackage{amsmath}
\usepackage{amsfonts}

\title{Hyperon Semileptonic decay constants with flavor SU(3) symmetry
breaking}

\author[a]{Ghil-Seok Yang,}
\author[b,c]{and Hyun-Chul Kim}

\affiliation[a]{Department of Physics, Soongsil University,\\
  Seoul 156-743, Republic of Korea}
\affiliation[b]{Department of Physics, Inha University,\\
  Incheon 402-751, Republic of Korea}
\affiliation[c]{School of Physics, Korea Institute for Advanced Study 
  (KIAS),\\ Seoul 130-722, Republic of Korea}

\emailAdd{ghsyang@ssu.ac.kr}
\emailAdd{hchkim@inha.ac.kr}

\abstract{
We investigate the hyperon semileptonic decay constants, $f_2/f_1$,
and $g_1/f_1$, within a general framework of a chiral soliton model.  
All relevant parameters for the SU(3) baryon wave functions were
unambiguously determined by using the experimental data for the masses
of the baryon octet and the decuplet. Using then the existing
experimental data for the magnetic moments of the baryon octet and the
decay constants of hyperon semileptonic decays, we are able to
determine all the hyperon semileptonic decay constants $f_2/f_1$ and
$g_1/f_1$ of the baryon octet unequivocally. In addition, we also
present the results of the axial-vector transition constants from 
the baryon decuplet to the octet.} 


\preprint{INHA-NTG-03/2015}

\begin{document}
\maketitle

\section{Introduction}
The Cabibbo-Kobayashi-Maskawa matrix elements $|V_{ud}|$ and
$|V_{us}|$ characterize the quark mixings in the process $d\to
ue^{-}\overline{\nu}_{e}$ and $s\to
ue^{-}\overline{\nu}_{e}$~\cite{Cabibbo1963,Kobayashi:1973fv} in the    
standard model. So far the most precise determination of $|V_{ud}|$
and $|V_{us}|$ are obtained respectively from super-allowed Fermi
transition together with pion decays, and from leptonic and
semileptonic kaon decays~\cite{Blucher:2005dc, Hardy:2013lga, 
  Antonelli:2010yf}. On the other hand, hyperon semileptonic decays
(HSDs) can also provide independent constraints on $|V_{ud}|$ and
$|V_{us}|$~\cite{Cabibbo:2003cu,Cabibbo:2003ea}. For last years, new
experimental results for HSD have been reported: the KTeV
collaboration first announced the measurement of the $\Xi^{0} 
\to \Sigma^{+} e^{-} \overline{\nu}_{e}$ decay~\cite{Affolder:1999pe}
and determined the corresponding form
factors~\cite{AlaviHarati:2001xk} as   
$f_{2}(0)/f_{1}(0)=2.0\pm1.2_{\textrm{stat}}\pm0.5_{\textrm{syst}}$
and $g_{1}(0)/f_{1}(0)=1.32_{-0.17\textrm{stat}}^{+0.21}\pm0.05_{\textrm{syst}}$.
The NA48/1 collaboration brought about the branching ratios for   
the same process with higher statistics in comparison with the KTeV
experiments: $g_{1}(0)/f_{1}(0)=1.21\pm0.05$ for 
$\Xi^{0} \to \Sigma^{+}e^{-} \overline{\nu}_{e}$~\cite{Batley:2006fc}.
Now there are experimental data for the $g_1(0)/f_1(0)$
ratios in five different decay channels in HSDs, collected by the
Particle Data Group (PDG)~\cite{Agashe:2014kda}. 

It is also of great importance to understand the HSD constants, since
they provide essential information on properties of the nucleon and
low-lying hyperons. The data for HSDs reveal experimentally the
pattern of flavor SU(3) symmetry breaking. In exact flavor SU(3)
symmetry the ratios of the axial-vector and vector constants $g_1/f_1$
are expressed only by the two constants $F$ and $D$. Similarly, 
the ratios of the vector constants $f_2/f_1$ are written in terms of
the anomalous magnetic moments of the proton and the neutron with
flavor SU(3) symmetry assumed. However, the experimental data for HSDs 
show that flavor SU(3) symmetry is manifestly broken. For example, the 
ratio $g_1/f_1=1.21\pm 0.05$ for the HSD process $\Xi^0\to \Sigma^+
e^- \overline{\nu}_e$ measured by the KTeV
collaboration~\cite{Affolder:1999pe,AlaviHarati:2001xk} is equal to
that of the 
neutron $\beta$ decay ($g_1/f_1=1.2695\pm0.0029$) in flavor SU(3)
symmetry. Thus, complete information on HSDs will furnish the pattern
of flavor SU(3) symmetry breaking in nature. It is also interesting to
see that the experimental data for the HSD constants give a clue how
isospin symmetry is broken. The above-mentioned ratio of the HSD
constants $g_1/f_1=1.20\pm 0.05$ from the KTeV collaboration should be 
equal to that of its isospin partner process $\Xi^{-} \to 
\Sigma^{0} e^{-} \overline{\nu}_e$ ($g_{1}(0) /
f_{1}(0) = 1.25_{-0.16}^{+0.14}$)~\cite{Bourquin:1983de} even with
flavor SU(3) symmetry breaking. If it is possible to measure more
precisely the HSD constants for the decay $\Xi^{-} \to \Sigma^{0}
e^{-} \overline{\nu}_e$, HSDs will shed light on how isospin symmetry 
is broken. 

For this reason, there has been a great deal of theoretical
works. Chiral perturbation theory ($\chi$PT) is used as a framework  
to analyze flavor SU(3) breaking effects on $f_1(0)$
beyond the first order~\cite{Villadoro:2006nj, Lacour:2007wm,
Faessler:2008ix,Guadagnoli:2006gj}, in which other form factors of
HSD have been also studied. The results for $V_{us}$ within HSDs 
were discussed in Ref.~\cite{Mateu:2005wi} and for the large
$N_{c}$ expansion in Ref.~\cite{FloresMendieta:2004sk}. 
In lattice QCD~\cite{Guadagnoli:2006gj} all form factors of the
$\Sigma^{-}\to nl\nu$ decay were first time studied and the results
were given as: 
$f_{2}(0)/f_{1}(0)=-1.52\pm0.81$ and $g_{1}(0) / f_{1}(0) = -0.287 \pm
0.052$. Flavor SU(3) symmetry breaking in HSDs was also investigated
within the $1/N_c$ expansion~\cite{FloresMendieta:1998ii}.  In the
chiral quark-soliton model ($\chi$QSM) with a 
\textit{model-independent approach} considered, the HSD constants 
were investigated many years ago~\cite{Kim:1999uf}. Being
distinguished from the self-consistent
$\chi$QSM~\cite{Kim:1997ts,Ledwig:2008ku}, almost all dynamical
parameters were fixed by using the experimental data of the masses and
the HSD constants for the baryon octet in this
\textit{model-independent} approach. In Ref.~\cite{Kim:1999uf},
however, the baryon wave functions were not completely determined and
experimental information was then not enough to fix unambiguously the 
parameters for the HSD constants. 
      
In the present work we will examine the HSD constants of the baryon
octet and decuplet within a general framework of a chiral soliton
model ($\chi$SM), fixing \textit{all} dynamical parameters
unequivocally. To derive the collective wave functions for the baryon
octet, one has to take into account both flavor SU(3) symmetry and
isospin symmetry breakings such that the experimental data for the
masses of the baryon octet can be employed as a whole. In particular,
sources of isospin symmetry breaking arise from the electromagnetic
(EM) interaction as well as from the mass difference of the up and
down quarks. In Ref.~\cite{Yang:2010id}, the EM mass differences
between baryons in the same isospin multiplet were analyzed within the
$\chi$SM and the corresponding model parameters were fixed by the
experimental data. Together with the isospin symmetry breaking from
the mass difference of the up and down quarks,
Ref.~\cite{Yang:2010fm,Yang:2011qe} showed that the collective wave
functions of the baryon octet and decuplet were uniquely
determined. In addition, a more complete analysis of the HSD  
constants can be carried out, the KTeV data for the $\Xi^0\to \Sigma^+ 
e^- \overline{\nu}_e$ process considered. Since now the five
experimental data for the ratios $g_1/f_1$, we are able to fix all 
dynamical parameters for the HSD axial-vector constants with the
singlet axial-vector charge from the data of deep inelastic
scattering employed. Similarly, all dynamical parameters
for the HSD vector constants can be fixed by using the
experimental data for the magnetic moments of the baryon octet. Using
the fixed parameters, we compute the HSD axial-vector and vector
constants for the baryon octet. In addition, we also present the
results of the transition axial-vector constants for the baryon
decuplet, which will be of great use in determining the meson-baryon
strong coupling constants.  

The present work is sketched as follows. In Section II, we explain 
how to fix all dynamical parameters, using the experimental data. We
also recapitulate relevant formulae for the HSD coupling constants. In
Section III, we show the results of the present work and discuss
physical implications of them. In the final Section, we summarize the
present work and draw conclusions.  Additional detailed formulae are
given in Appendices. 
\section{Baryon matrix elements of the vector and axial-vector 
  currents}    
The transition matrix elements of the vector and axial-vector
currents for the baryon octet are expressed respectively in terms of
the six real form factors:   
\begin{eqnarray}
\langle B_{2}|V_{\mu}^{\chi}|B_{1}\rangle 
&=& 
\bar{u}_{B_{2}}(p_{2},s_2)
\left[f_{1}^{B_1\to B_2}(q^{2})\gamma_{\mu} - \frac{if_{2}^{B_1\to B_2} 
    (q^{2}) \sigma_{\mu\nu}q^{\nu}}{M_{B_1}}  \right. \cr
&& \left.\hspace{5cm} +\; \frac{f_{3}^{B_1\to B_2}
  (q^{2}) q_{\mu}}{M_{B_1}}\right] u_{B_{1}}(p_{1},s_1),  
\cr
\langle B_{2}|A_{\mu}^{\chi}|B_{1}\rangle 
&=& 
\bar{u}_{B_{2}}(p_{2},s_2)
\left[g_{1}^{B_1\to B_2}(q^{2})\gamma_{\mu} + \frac{ig_{2}^{B_1\to B_2} 
    (q^{2}) \sigma_{\mu\nu}q^{\nu}}{M_{B_1}} \right. \cr
&& \left.\hspace{5cm} +\; \frac{g_{3}^{B_1\to B_2}   (q^{2})
   q_{\mu}}{M_{B_1}}\right] 
\gamma_{5}u_{B_{1}}(p_{1},s_1),   
\label{eq:2}
\end{eqnarray}
where the vector and axial-vector currents are defined as  
\begin{equation}
V_{\mu}^{\chi} (x) 
\;=\;
\bar{\psi}(x)\gamma_{\mu}\frac{1}{2}
\lambda^{\chi} \psi(x),\;\;\;\;
A_{\mu}^{\chi} (x) 
\;=\;
\bar{\psi}(x)\gamma_{\mu}\gamma_{5}\frac{1}{2} 
\lambda^{\chi} \psi(x).  
\label{eq:3}
\end{equation}
The $u_{B_1}\,(\bar{u}_{B_2}) $ denote the Dirac spinors corresponding
to the initial and final baryon states, respectively. The
$\lambda^\chi$ designate flavor Gell-Mann matrices for 
strangeness conserving $\Delta S=0$ transitions
($\chi\,=\,3,\,8,\,1\,\pm\, i\,2$) and for  
$|\Delta S|=1$ ones ($\chi\,=\,4\,\pm\, i\,5$), respectively. The
$q^{2}=-Q^{2}$ stands for the square of the momentum transfer
$q=p_{2}-p_{1}$. The form factors $f_{i}^{B_1\to B_2}$ and 
$g_{i}$ are real quantities due to $CP$-invariance, depending only on
the square of the momentum transfer. We can neglect $f_3 ^{B_1\to
  B_2}$ and $g_{3}^{B_1\to B_2}$ for the reason that their
contributions to the decay rate is proportional to the ratio
$m_{l}^{2}/M_{B_1}^{2}\ll 1$, where $m_{l}$ represents a mass
of the lepton ($e$ or $\mu$) in the final state and that of the baryon
in the initial state, $M_{B_1}$, respectively. The $g_2^{B_1\to B_2}$
are finite only with the effects of flavor $\mathrm{SU}(3)$ symmetry   
and isospin symmetry breakings because of its opposite $G$ parity to
the axial-vector current, so that they are very small for the baryon
octet.  Moreover, the Ademollo-Gatto theorem~\cite{Ademollo:1964sr}
does not allow $f_1^{B_2\to B_1}$ to acquire the linear-order
corrections of flavor $\mathrm{SU}(3)$ symmetry breaking, so that
$f_1^{B_2\to B_1}$ are merely expressed in terms of the SU(3)
Clebsch-Gordan coefficients. 

Concerning the HSD constants for the transition from the baryon
decuplet to the octet, it is difficult to determine the vector HSD 
constants because of a lack of experimental data for the $M1$ and $E2$ 
transitions from the baryon decuplet to the octet.  
For this reason, we will consider only the axial-vector transitions in
the present work. There are several different ways of decomposing the
matrix elements of the axial-vector current between the baryon
decuplet and the octet. We will follow here the formalism of the Alder
form factors 
$C_{i}^{A}\left(q^{2}\right)$~\cite{Llewellyn
  Smith:1971zm,Hemmert:1994ky,Golli:2002wy,Alexandrou:2006mc} 
to describe the axial-vector transitions of the baryon decuplet.
The matrix elements between the baryon decuplet and the octet can be
written as follows
\begin{eqnarray}
\left\langle B_{8}\left|A_{\mu}^{\chi}\right|B_{10}\right\rangle  
& = & 
\overline{u}_{B_{8}}(p_{2},\, s_{2})\left[C_{5}^{A}(q^{2})g_{\mu\nu}
+ C_{6}^{A}(q^{2})q_{\mu}q_{\nu}\right.\cr
&  & \left.+\left\{
    C_{3}^{A}(q^{2})\gamma^{\alpha}
  + C_{4}^{A}(q^{2})p^{\prime\alpha}\right\} 
\left(q_{\alpha}g_{\mu\nu}-q_{\nu}g_{\alpha\mu}\right)\right]
u_{B_{10}}^{\nu}(p_{1},\; s_{1}),
\label{eq:Adler}
\end{eqnarray}
where the $u_{B_{10}}^{\nu}\left(p_{1},\, s_{1}\right)$ represents the 
Rarita-Schwinger spinor that describes the baryon decuplet with spin
$3/2$.  $p'$ is defined as $p'=(M_{B_{10}},0,0,0)$.

Since we want to determine the HSD constants within the framework of a 
$\chi$SM with all dynamical parameters in the model fixed by the
experimental data, we first explain the $\chi$SM in brief. The
detailed formalism can be found for example in Ref.~\cite{Yang:2010fm}
and references therein. The $\chi$SM is characterized by the following
collective Hamiltonian:  
\begin{equation}
  \label{eq:CollHamil}
H= M_{\mathrm{cl}} + H_{\mathrm{rot}} + H_{\mathrm{sb}},    
\end{equation}
where $M_{\mathrm{cl}}$ stands for the classical soliton mass. The
collective Hamiltonian $H_{\mathrm{rot}}$ comes from the collective
quantization of the chiral soliton by considering its slow
rotation. We often call it the $1/N_c$ rotational correction and
express it as 
\begin{equation}
H_{\mathrm{rot}}\;=\; \frac1{2I_1} \sum_{i=1}^3 \hat{J}_i^2 +
\frac1{2I_2} \sum_{p=4}^7 \hat{J}_p^2,   
\end{equation}
where $I_{1,2}$ denote the moments of inertia of the soliton, which
depend on dynamics of specific formulations of the
$\chi$SM. $\hat{J}_{i}$ stand for the usual spin operators and $\hat{J}_p$
represent the group generators corresponding to the right rotation in
flavor SU(3) space. The last term in eq.(\ref{eq:CollHamil}) comes
from the flavor SU(3) symmetry breaking, which are expressed as 
\begin{eqnarray}
H_{\mathrm{sb}} &=& (m_d-m_u) \left(\frac{\sqrt{3}}{2} \alpha
  D_{38}^{(8)} (\mathcal{R}) + \beta \hat{T}_3 +\frac12 \gamma
  \sum_{i=1}^3 D_{3i}^{(8)} (\mathcal{R}) \hat{J}_i\right) \cr
& & + \;(m_s - \overline{m}) \left(\alpha D_{88}^{(8)} (\mathcal{R}) +
    \beta \hat{Y} + \frac1{\sqrt{3}} \gamma \sum_{i=1}^3 D_{8i}^{(8)}
    (\mathcal{R}) \hat{J}_i\right) \cr
&& + \;(m_u + m_d + m_s) \sigma,
  \label{eq:SU3BR}
\end{eqnarray}
where $m_{u}$, $m_d$, and $m_s$ denote the up, down, and strange
current quark masses, respectively. $\overline{m}$ designates the
average of the up and down current quark
mass. $D_{ab}^{(R)}(\mathcal{R})$ are the SU(3) Wigner $D$ functions
in a $R$ irreducible representation. $\hat{T}_3$ and $\hat{Y}$ stand
for the operators corresponding to the third component of the isospin
and the hypercharge, respectively. 
The dynamical parameters $\alpha$, $\beta$, and $\gamma$ encode
a specific dynamics of a certain chiral soliton model. They are
defined as 
\begin{equation}
  \label{eq:AlphaBetaGamma}
\alpha \;=\; -\left(\frac23 \frac{\Sigma_{\pi N}}{m_u+m_d} -
  \frac{K_2}{I_2}\right),\;\;\;\; \beta \;=\;
-\frac{K_2}{I_2},\;\;\;\; 
\gamma \;=\; 2\left(\frac{K_1}{I_1}  - \frac{K_2}{I_2} \right),
\end{equation}
where $\Sigma_{\pi N}$ indicates the $\pi N$ sigma term. The $\sigma$
in eq.~(\ref{eq:SU3BR}) is defined as 
\begin{equation}
\sigma \;=\; -(\alpha+\beta) \;=\; \frac13 \frac{\Sigma_{\pi
    N}}{\overline{m}}.
\end{equation}

In the $\chi$SM, the value of the eighth component of the soliton
angular velocity $J_8$ provides a very important constraint on the
collective quantization. In the Skyrme model, it has its origin in 
the Wess-Zumino term~\cite{Witten:1983tx,Guadagnini:1983uv,
  Jain:1984gp}, whereas it is related to the presence of the 
discrete valence quark level in the Dirac-sea spectrum in the SU(3)
chiral quark-soliton model~\cite{Blotz:1992pw, Christov:1995vm}.
Its presence has no effects on the chiral soliton but allows one to 
consider only the flavor SU(3) irreducible representations with zero
triality. 

The wave functions for the baryon octet and decuplet are derived by
diagonalizing the collective SU(3)
Hamiltonian of the $\chi$SM \cite{Blotz:1992pw, Christov:1995vm,
  Yang:2010fm, Yang:2011qe} and are expressed in terms of the SU(3)
Wigner $D$ functions:
\begin{equation}
\langle A|\mathcal{R},\, B(Y\, T\, T_{3},\; Y^{\prime}\, J\,
J_{3})\rangle 
\;=\; \sqrt{\textrm{dim}(\mathcal{R})}\,(-)^{J_{3}
+Y^{\prime}/2}\, D_{(Y,\, T,\, T_{3})(-Y^{\prime},\,
J,\,-J_{3})}^{(\mathcal{R})*}(A),
\label{eq:5}  
\end{equation}
where $\mathcal{R}$ designates the allowed irreducible representations 
of the flavor $\mathrm{SU}(3)$ group, i.e. $\mathcal{R} \, = \,8, \,10, \,
\cdots$. $Y$ $T$, and $T_{3}$ denote the corresponding hypercharge, 
isospin and its third component, respectively. The 
right hypercharge is constrained to be
$Y^{\prime}=2J_8/\sqrt{3}=-N_c/3=-1$ that selects a tower of allowed
flavor SU(3) representations. The baryon octet and decuplet, 
which are the lowest representations among them, coincide with
those of the quark model. 

In the presence of flavor $\mathrm{SU}(3)$ symmetry breaking,
a baryon state is no more pure state but a mixed one with higher
representations. Taking into account the strange current quark mass
$m_{\mathrm{s}}$ as a perturbation, we obtain the collective wave
functions for the baryon octet and decuplet mixed with higher
representations as  
\begin{eqnarray}
\left|B_{8}\right\rangle  & = & \left|8_{1/2},B\right\rangle \;+\;
c_{\overline{10}}^{B}\left|\overline{10}_{1/2},B\right\rangle \;+\;
c_{27}^{B}\left|27_{1/2},B\right\rangle ,\cr 
\left|B_{10}\right\rangle  & = & \left|10_{3/2},B\right\rangle \;+\;
a_{27}^{B}\left|27_{3/2},B\right\rangle \;+\;
a_{35}^{B}\left|35_{3/2},B\right\rangle,
\label{eq:9}
\end{eqnarray}
where the spin indices $J_{3}$ have been suppressed. The
$m_{\mathrm{s}}$-dependent coefficients in eq.~(\ref{eq:9}) are
written as
\begin{eqnarray}
c_{\overline{10}}^{B} 
& = &
c_{\overline{10}}\left[\kern-0.5em 
\begin{array}{c} 
\sqrt{5}\\
0\\
\sqrt{5}\\
0
\end{array}\kern-0.2em \right]\kern-0.2em ,\;
c_{27}^{B}=c_{27}\left[\kern-0.5em 
\begin{array}{c} 
\sqrt{6}\\
3\\
2\\
\sqrt{6}
\end{array}\kern-0.2em \right]\kern-0.2em ,\;
a_{27}^{B}=a_{27}\left[\kern-0.5em 
\begin{array}{c} 
\sqrt{15/2}\\
2\\
\sqrt{3/2}\\
0
\end{array}\kern-0.2em \right]\kern-0.2em ,\;
a_{35}^{B}=a_{35}\left[\kern-0.5em 
\begin{array}{c} 
5/\sqrt{14}\\
2\sqrt{5/7}\\
3\sqrt{5/14}\\
2\sqrt{5/7}
\end{array}\kern-0.2em \right],
\label{eq:10}
\end{eqnarray}
respectively in the basis $[N,\;\Lambda,\;\Sigma,\;\Xi]$ and
$[\Delta,\;\Sigma^{\ast},\;\Xi^{\ast},\;\Omega]$. 
The mixing coefficients in eq.~(\ref{eq:10}) that contain
$m_{\mathrm{s}}$ and $\overline{m}$ are written as  
\begin{eqnarray}
c_{\overline{10}} 
& = &
-\frac{I_{2}}{15}\left(m_{s}-\overline{m}\right)
\left(\alpha+\frac{1}{2}\gamma\right),\;\;\;\;\; 
c_{27} = -\frac{I_{2}}{25}\left(m_{s}-\overline{m}\right)
\left(\alpha-\frac{1}{6}\gamma\right),\cr  
a_{27} 
& = & 
-\frac{I_{2}}{8}\left(m_{s}-
  \overline{m}\right)\left(\alpha+\frac{5}{6}\gamma\right),\;\;\;\;\; 
a_{35} =
-\frac{I_{2}}{24}\left(m_{s}-\overline{m}\right)
\left(\alpha-\frac{1}{2}\gamma\right).
\label{eq:12} 
\end{eqnarray}

In order to determine the dynamical parameters $\alpha$, $\beta$,
$\gamma$ and the mixing coefficients in eq.(\ref{eq:12}) by using the
experimental data, we have to take into account the effects of isospin
symmetry breaking, which consist of the electromagnetic interactions
and the up and down quark mass differences. The corresponding
formalisms and analyses in detail can be found in
Ref.~\cite{Yang:2010id} and Ref.~\cite{Yang:2010fm}, respectively.   
The numerical values of the coefficients
that were obtained from Ref.~\cite{Yang:2010fm} are given as
\begin{eqnarray}
c_{\overline{10}} &=& 
0.0434 \pm 0.0006,\;\;\;
c_{27} \;=\; 0.0203\pm
0.0003,\cr
a_{27} &=& 0.0903\pm 0.0013,\;\;\; 
a_{35} \;=\; 0.0181\pm 0.0003.
  \label{eq:13}
\end{eqnarray}
Note that it is of great importance to know these mixing coefficients,
since they allow one to carry out an unambiguous analysis of the
vector and axial-vector HSD constants. 

In a $\chi$SM, the collective operators for the HSD vector and
axial-vector constants can be expressed in terms of the SU(3)
Wigner $D$ functions~\cite{Kim:1997ip, Kim:1999uf,Yang:2004jr}:
\begin{eqnarray}
\hat{f}_{2}^{B_{1}\rightarrow B_{2}} 
& = & 
w_{1}\, D_{\chi 3}^{(8)} 
+ 
w_{2}d_{pq3}\, D_{\chi p}^{(8)}\,\hat{J}_{q} 
+ 
\frac{w_{3}}{\sqrt{3}}\,D_{\chi 8}^{(8)}\,\hat{J}_{3}
+\frac{w_{4}}{\sqrt{3}}\, d_{pq3}D_{\chi p}^{(8)}\, D_{8q}^{(8)}  \cr
&& 
+\; w_{5}\,\left(D_{\chi 3}^{(8)}\, D_{88}^{(8)} + D_{\chi 8}^{(8)}\, 
D_{83}^{(8)}\right)
+w_{6}\,\left(D_{\chi 3}^{(8)}\,
D_{88}^{(8)}-D_{\chi 8}^{(8)}\, D_{83}^{(8)}\right),\cr
\hat{g}_{1}^{B_{1}\rightarrow B_{2}} 
& = & 
a_{1}\, D_{\chi 3}^{(8)}
+ 
a_{2}d_{pq3}
\, D_{\chi p}^{(8)}\,\hat{J}_{q}
+
\frac{a_{3}}{\sqrt{3}}\,D_{\chi 8}^{(8)}\,\hat{J}_{3}
+\frac{a_{4}}{\sqrt{3}}\, d_{pq3}D_{\chi p}^{(8)}\, 
D_{8q}^{(8)}
 \cr
&& 
+\; a_{5}\,\left(D_{\chi 3}^{(8)}\, D_{88}^{(8)}+D_{\chi 8}^{(8)}\, 
D_{83}^{(8)}\right)
+ a_{6}\,\left(D_{\chi 3}^{(8)}\,
D_{88}^{(8)}-D_{\chi 8}^{(8)}\, D_{83}^{(8)}\right),
\label{eq:14}
\end{eqnarray}
where $w_i$ and $a_{i}$ stand for the parameters encoding
the specific dynamics of a chiral solitonic model as in the case of
the $\alpha$, $\beta$, and $\gamma$. For example, they
can be explicitly computed by using the $\chi$QSM~\cite{Kim:1997ts,  
  Ledwig:2008ku}. Each $w_i$ has its own dynamical meaning:
$w_1\,(a_1)$ corresponds to the leading-order contribution,
$w_2\,(a_2)$ and $w_3\,(a_3)$ arise from the rotational $1/N_c$
corrections, and $w_4\,(a_4)$, $w_5\,(a_5)$ and $w_6\,(a_6)$ are
originated from flavor $\mathrm{SU}(3)$ symmetry breaking, in which
the strange current quark mass $m_{\mathrm{s}}$ is included. There are
two different linear $m_s$ corrections to the HSD constants
$f_2^{B_1\to   B_2}$ and $g_1^{B_1\to B_2}$: The $m_s$ corrections
from the collective operators given in eq.(\ref{eq:14}) and those from
the baryon wave functions because of the mixing coefficients in
eq.(\ref{eq:12}).  

\section{Analysis of the HSD constants for the baryon octet}
We are now in a position to determine the parameters $w_i$ and $a_i$
in eq.(\ref{eq:14}). The experimental data for the magnetic moments of
the baryon octet, which are listed in Table~\ref{tab:MagInput1}, can be  used 
for the determination of $w_i$.  
\begin{table}[htp]
  \begin{tabular}{ll}
\hline \hline
    & Experimental data~\cite{Agashe:2014kda} 
\\
\hline 
 $\mu_p$ 
&$\phantom{-}2.792847356 \pm 0.000000023$ \\
 $\mu_n$ 
& $-1.9130427 \pm 0.0000005$ \\
 $\mu_\Lambda$ 
& $-0.613\pm 0.004$ \\
 $\mu_{\Sigma^+}$ 
& $\phantom{-}2.458 \pm0.010$ \\
$\mu_{\Sigma^-}$ 
& $-1.160 \pm 0.025$ \\
$\mu_{\Xi^0}$ 
& $-1.250 \pm 0.014$ \\ 
$\mu_{\Xi^+}$ 
& $-0.6507 \pm 0.0025$ \\
\hline \hline
\end{tabular}
\caption{The experimental data for the magnetic moments of the baryon
  octet~\cite{Agashe:2014kda} in units of the nuclear magneton
  $\mu_N$, which are used as input in the present work.}  
\label{tab:MagInput1}
\end{table}

In fact, the analyses of the magnetic moments in a similar scheme to 
the present one were carried out in Refs.~\cite{Kim:1997ip,
  Kim:1998gt, Yang:2004jr}. However, it was then not possible to
determine all relevant parameters uniquely, so that the results were
shown as functions either of $m_{s}$ or of the $\pi N$ sigma term. 
With the mixing coefficients in eq.(\ref{eq:13}) completely
determined, however, we are able to proceed to fix
the parameters $w_i$ unambiguously by using the experimental data for
the baryon magnetic moments listed in Table~\ref{tab:MagInput1}. Since
we have the seven independent experimental data as shown in 
Table~\ref{tab:MagInput1}, whereas we need to find the six parameters
($w_i$), we use the $\chi^2$-fit method. Let us set the magnetic
moments of the octet baryon and the parameters $w_i$ as vectors 
\begin{equation}
\bm \mu 
\;=\; 
(\mu_p,\,\mu_n,\,\mu_\Lambda,\,\mu_{\Sigma^+},
\mu_{\Sigma^-},\,\mu_{\Xi^0},\,\mu_{\Xi^-}),\;\;\;\; 
\bm w 
\;=\; (w_1,\,w_2,\,w_3,\,w_4,\,w_5,\,w_6)  
\end{equation}
such that the relation between $\bm \mu$ and $\bm \omega$ is cast
into a matrix equation 
\begin{equation}
\bm \mu \;=\; \bm A \cdot \bm w,  
\label{eq:MatrixEq16}
\end{equation}
where
\begin{equation}
\bm A\; = \;
\begin{bmatrix}
-\frac2{15} - \frac{4}{45} c_{27}
& \frac1{15} - \frac{8 }{45}c_{27} 
& 
\frac1{30} + \frac{2}{15}c_{27}
& -\frac2{135} 
& - \frac1{18} 
&
-\frac1{30} \\[0.3em]   
-\frac1{10} + \frac{1}{3}c_{\overline{10}} -\frac{2 }{45}c_{27} 
&
-\frac1{20} + \frac{1}{3}c_{\overline{10}} - \frac{4}{45} c_{27}
&  
\frac1{60}  + \frac{1}{6}c_{\overline{10}} + \frac{1}{15}c_{27} 
& \frac7{270} 
& \frac1{18} 
& \frac1{30} 
\\[0.3em] 
   \frac1{20} - \frac{1}{10} c_{27} 
& -\frac1{40} - \frac{1}{5}  c_{27}
& 
\frac1{120} + \frac{3 }{20} c_{27}
& -\frac1{60} 
& 0 
& 0 
\\[0.3em] 
 -\frac2{15} - \frac{2 }{45} c_{27}
& \frac1{15} - \frac{4}{45} c_{27}
& \frac1{30} + \frac{1}{15}  c_{27}
& -\frac1{135} 
& -\frac1{90} 
& \frac1{30} 
\\[0.3em] 
   \frac1{30} + \frac{1}{3}c_{\overline{10}} - \frac{2}{45} c_{27} 
& -\frac1{60} + \frac{1}{3}c_{\overline{10}} - \frac{4}{45} c_{27} 
& -\frac1{20} + \frac{1}{6}c_{\overline{10}} + \frac{1}{15} c_{27}
& \frac7{270} 
& \frac1{18} 
& -\frac1{30} 
\\[0.3em] 
   \frac1{10} - \frac{2}{45} c_{27}
& -\frac1{20} - \frac{4}{45}c_{27} 
&  \frac1{60} + \frac{1}{15} c_{27}
& -\frac1{135} 
& - \frac1{90} 
& -\frac1{30} 
\\[0.3em] 
   \frac1{30} - \frac{4}{45} c_{27}  
& -\frac1{60} - \frac{8}{45}  c_{27}
& -\frac1{20} + \frac{2}{15} c_{27} 
& -\frac2{135} 
& - \frac1{18} 
&
\frac1{30}  
\end{bmatrix}.
\end{equation}
Having solved eq.(\ref{eq:MatrixEq16}), we determine the parameters 
$w_i$ as listed in Table~\ref{tab:w}.
\begin{table}[htp]
\begin{tabular}{cccccc}
\hline \hline
$w_{1}$ 
&   
$w_{2}$ 
&   
$w_{3}$ 
&   
$w_{4}$ 
&   
$w_{5}$ 
&   $w_{6}$
\tabularnewline\hline
$-13.51\pm0.01$  
&   
$4.15\pm0.93$  
&   
$8.54\pm0.86$  
&   
$-3.79\pm0.21$  
&   
$-4.93\pm0.86$  
&   
$-2.01\pm0.84$
\\
\hline \hline
\end{tabular}
\caption{The dynamical parameters $w_i$ for the HSD vector constants
  $f_2$} 
\label{tab:w} 
\end{table}
We want to point out that the results of $w_i$ in
Ref.~\cite{Kim:1997ip} suffered from an uncertainty, so that the
magnetic moments of the baryon decuplet were given as a function of an
unknown parameter $p$. On the other hand,   
the results of $w_i$ in the present work are uniquely determined, as
shown in Table~\ref{tab:w}. 

Using the results of $w_i$ in Table~\ref{tab:w}, we obtain
straightforwardly $f_2/f_1$ for various HSD modes:
\begin{eqnarray}
\left[\begin{array}{c}
f_{2}/f_{1}\left(n\rightarrow p\right)\\
f_{2}/f_{1}\left(\Sigma^{+}\rightarrow\Lambda\right)\\
f_{2}/f_{1}\left(\Sigma^{-}\rightarrow n\right)\\
f_{2}/f_{1}\left(\Xi^{-}\rightarrow\Lambda\right)\\
f_{2}/f_{1}\left(\Xi^{-}\rightarrow\Sigma^{0}\right)\\
f_{2}/f_{1}\left(\Sigma^{-}\rightarrow\Lambda\right)\\
f_{2}/f_{1}\left(\Lambda\rightarrow p\right)\\
f_{2}/f_{1}\left(\Sigma^{-}\rightarrow\Sigma^{0}\right)\\
f_{2}/f_{1}\left(\Xi^{-}\rightarrow\Xi^{0}\right)\\
f_{2}/f_{1}\left(\Xi^{0}\rightarrow\Sigma^{+}\right)\\
\end{array}\right] 
& = & 
\mathbf{M}_V\;
\left[\begin{array}{c}
w_{1}\\
w_{2}\\
w_{3}\\
w_{4}\\
w_{5}\\
w_{6}
\end{array}\right]
\label{eq:g1f1M}
\end{eqnarray}
with the matrix $\mathbf{M}_V$ expressed as 
\begin{equation}
 \mathbf{M}_V
\;=\; 
\begin{bmatrix}
-\frac{7}{30} - \frac{1}{3}c_{\overline{10}} - \frac{2}{45}c_{27} 
& \frac{7}{60} - \frac{1}{3}c_{\overline{10}} - \frac{4}{45}c_{27} 
& \frac{1}{60} - \frac{1}{6}c_{\overline{10}} + \frac{1}{15}c_{27} 
& -\frac{11}{270} 
& -\frac{1}{9} 
& -\frac{1}{15}\\[0.3em]
-\frac{3}{20} - \frac{1}{4}c_{\overline{10}} - \frac{1}{12}c_{27} 
& \frac{3}{40} - \frac{1}{4}c_{\overline{10}} - \frac{1}{6}c_{27} 
& -\frac{1}{40} - \frac{1}{8}c_{\overline{10}} + \frac{1}{8}c_{27} 
& -\frac{7}{180} 
& -\frac{1}{30} 
& 0\\[0.3em]
\frac{1}{15} - \frac{1}{45}c_{27} 
& -\frac{1}{30} - \frac{2}{45}c_{27} 
& \frac{1}{15} + \frac{1}{30}c_{27} 
& -\frac{1}{270} 
& -\frac{1}{45} 
& \frac{1}{30}\\[0.3em]
-\frac{1}{30} - \frac{1}{30}c_{27} 
& \frac{1}{60} - \frac{1}{15}c_{27} 
& \frac{1}{20} + \frac{1}{20}c_{27} 
& -\frac{1}{180} 
& \frac{1}{30} 
& -\frac{1}{30}\\[0.3em]
-\frac{7}{30} + \frac{1}{6}c_{\overline{10}} + \frac{1}{45}c_{27} 
& \frac{7}{60} + \frac{1}{6}c_{\overline{10}} + \frac{2}{45}c_{27} 
& \frac{1}{60} + \frac{1}{12}c_{\overline{10}} - \frac{1}{30}c_{27} 
& \frac{11}{540} 
& \frac{1}{18} 
& \frac{1}{30}\\[0.3em]
-\frac{3}{20} - \frac{1}{4}c_{\overline{10}} - \frac{1}{12}c_{27} 
& \frac{3}{40} - \frac{1}{4}c_{\overline{10}} - \frac{1}{6}c_{27} 
& -\frac{1}{40} - \frac{1}{8}c_{\overline{10}} + \frac{1}{8}c_{27} 
& -\frac{7}{180} 
& -\frac{1}{30} 
& 0\\[0.3em]
-\frac{2}{15} + \frac{1}{6}c_{\overline{10}} + \frac{1}{30}c_{27} 
& \frac{1}{15} + \frac{1}{6}c_{\overline{10}} + \frac{1}{15}c_{27} 
& \frac{1}{30} + \frac{1}{12}c_{\overline{10}} - \frac{1}{20}c_{27} 
& \frac{1}{45} 
& 0 
& -\frac{1}{30}\\[0.3em]
-\frac{1}{12} - \frac{1}{6}c_{\overline{10}} 
&
\frac{1}{24} - \frac{1}{6}c_{\overline{10}} 
& \frac{1}{24} - \frac{1}{12}c_{\overline{10}} 
& -\frac{1}{60} 
& -\frac{1}{30} 
& \frac{1}{30}\\[0.3em]
\frac{1}{15} + \frac{2}{45}c_{27} 
& -\frac{1}{30} + \frac{4}{45}c_{27} 
& \frac{1}{15} - \frac{1}{15}c_{27} 
& \frac{1}{135} 
& \frac{2}{45} 
& -\frac{1}{15}\\[0.3em]
-\frac{7}{30} + \frac{1}{6}c_{\overline{10}} + \frac{1}{45}c_{27} 
& \frac{7}{60} + \frac{1}{6}c_{\overline{10}} + \frac{2}{45}c_{27} 
& \frac{1}{60} + \frac{1}{12}c_{\overline{10}} - \frac{1}{30}c_{27} 
& \frac{11}{540} 
& \frac{1}{18} 
& \frac{1}{30}
\end{bmatrix}
.
\label{eq:matrixMV}
\end{equation}

It is well known that when the effects of SU(3) symmetry breaking are
turned off, the vector HSD constants $f_2(0)/f_1(0)$ are expressed in
terms of the anomalous magnetic moments of the proton and the neutron
as listed in Table~\ref{tab:Sym3}. Note that the constants $f_1(0)$ do
not have any linear $m_{\mathrm{s}}$ corrections because of the
Ademollo-Gatto theorem~\cite{Ademollo:1964sr}. 
\begin{table}[htp]
\begin{tabular}{lccc}
\hline \hline 
Decay mode  
& $f_{1}(0)$  
& $f_{2}(0)$  
& $f_{2}(0)/f_{1}(0)$  
\\[0.3em]
\hline 
$n\rightarrow p$  
& $1$  
& $\frac{1}{2}\left(\kappa_{p} - \kappa_{n}\right)$  
& $\frac{1}{2}\left(\kappa_{p} - \kappa_{n}\right)$  
\\[0.3em]
$\Sigma^{-}\rightarrow\Lambda$  
& $0$  
& $-\frac{1}{2}\sqrt{\frac{3}{2}}\kappa_{n}$  
& $\sqrt{\frac{3}{2}}f_{2}^{\,\Sigma^{-}\rightarrow\Lambda}(0)$  
\\[0.3em]
$\Sigma^{0}\rightarrow\Sigma^{+}$  
& $\sqrt{2}$  
& $\frac{1}{\sqrt{2}}\left(\kappa_{p} + \frac{1}{2}\kappa_{n}\right)$  
& $\frac{1}{2}\left(\kappa_{p} + \frac{1}{2}\kappa_{n}\right)$  
\\[0.3em]
$\Sigma^{-}\rightarrow\Sigma^{0}$  
& $\sqrt{2}$  
& $\frac{1}{\sqrt{2}}\left(\kappa_{p} + \frac{1}{2}\kappa_{n}\right)$  
& $\frac{1}{2}\left(\kappa_{p} + \frac{1}{2}\kappa_{n}\right)$  
\\[0.3em]
$\Xi^{-}\rightarrow\Xi^{0}$ 
& $1$  
& $\frac{1}{2}\left(\kappa_{p} + 2\kappa_{n}\right)$  
& $\frac{1}{2}\left(\kappa_{p} + 2\kappa_{n}\right)$  
\\[0.3em]
\hline 
$\Lambda\rightarrow p$  
& $\sqrt{\frac{3}{2}}$  
& $\frac{1}{2}\sqrt{\frac{3}{2}}\kappa_{p}$  
& $\frac{1}{2}\kappa_{p}$  
\\[0.3em]
$\Sigma^{0}\rightarrow p$  
& $\frac{1}{\sqrt{2}}$  
& $\frac{1}{2\sqrt{2}}\left(\kappa_{p} + 2\kappa_{n}\right)$  
& $\frac{1}{2}\left(\kappa_{p} + 2\kappa_{n}\right)$  
\\[0.3em]
$\Sigma^{-}\rightarrow n$  
& $1$  
& $\frac{1}{2}\left(\kappa_{p} + 2\kappa_{n}\right)$  
& $\frac{1}{2}\left(\kappa_{p} + 2\kappa_{n}\right)$  
\\[0.3em]
$\Xi^{0}\rightarrow\Sigma^{+}$  
& $1$  
& $\frac{1}{2}\left(\kappa_{p} - \kappa_{n}\right)$  
& $\frac{1}{2}\left(\kappa_{p} - \kappa_{n}\right)$  
\\[0.3em]
$\Xi^{-}\rightarrow\Lambda$  
& $\sqrt{\frac{3}{2}}$  
& $\frac{1}{2}\sqrt{\frac{3}{2}}\left(\kappa_{p} + \kappa_{n}\right)$  
& $\frac{1}{2}\left(\kappa_{p} + \kappa_{n}\right)$  
\\[0.3em]
$\Xi^{-}\rightarrow\Sigma^{0}$  
& $\frac{1}{\sqrt{2}}$  
& $\frac{1}{2\sqrt{2}}\left(\kappa_{p} - \kappa_{n}\right)$  
& $\frac{1}{2}\left(\kappa_{p} - \kappa_{n}\right)$  
\\[0.3em]
\hline 
\hline
\end{tabular}
\caption{The expressions of $f_{1}(0)$, $f_{2}(0)$, and $g_{1}(0)$ in
  exact SU(3) symmetry. The $\kappa_{p}$ and the $\kappa_{n}$ denote
  the anomalous magnetic moments of the proton and the neutron,
  respectively.} 
\label{tab:Sym3}
\end{table}
However, one has to use the values of $\kappa_p$ and $\kappa_n$ in the
exact SU(3) symmetry to compute $f_2(0)$ in SU(3) symmetry, since the
experimental data for the nucleon anomalous magnetic moments already
include the effects of SU(3) symmetry breaking. 
The numerical values of anomalous magnetic moments of proton and
neutrons for the contributions of the order $\mathcal{O}(m_{s}^{0})$
and $\mathcal{O}(m_{s}^{0})+\mathcal{O}(m_{s}^{1})$ are obtained as
follows:  
\begin{eqnarray}
\kappa_{p}^{\mathrm{sym}}\left[\mathcal{O}(m_{s}^{0})\right] 
& = & 
1.363\pm0.069,\cr
\kappa_{n}^{\mathrm{sym}}\left[\mathcal{O}(m_{s}^{0})\right] 
& = & 
-1.416\pm0.049,\cr
\kappa_{p}\left[\mathcal{O}(m_{s}^{0})
+ \mathcal{O}(m_{s}^{1})\right] 
& = & 
1.793\pm0.087,\cr
\kappa_{n}\left[\mathcal{O}(m_{s}^{0})
+ \mathcal{O}(m_{s}^{1})\right] 
& = & 
-1.913\pm0.069,
\label{eq:anomalMag}
\end{eqnarray}
where $\kappa_{p}^{\mathrm{sym}}$ and $\kappa_{n}^{\mathrm{sym}}$
represent the anomalous magnetic moments of the proton and the neutron
in the exact SU(3) symmetry. We can estimate the effect of SU(3)
symmetry breaking from eq.(\ref{eq:anomalMag}), which is approximately
$25\,\%$. 

\begin{table}[htp] 
\begin{tabular}{lrrr}
\hline \hline
Decay mode  
& $(f_{2}/f_{1})^{\mathrm{sym}}$
&  $(f_{2}/f_{1})^{\mathrm{br}}$\phantom{s}
& Experimental data~\cite{Agashe:2014kda} 
\tabularnewline
\hline 
$n\rightarrow p$  
& $1.389\pm0.042$  
& $1.853\pm0.056$  
& $ $ 
\tabularnewline
$\Sigma^{-}\rightarrow\Lambda$  
& $1.062\pm0.037$  
& $1.435\pm0.052$  
 $ $ 
\tabularnewline
$\Sigma^{-}\rightarrow\Sigma^{0}$  
& $0.328\pm0.037$  
& $0.418\pm0.047$  
& $ $ 
\tabularnewline
$\Xi^{-}\rightarrow\Xi^{0}$  
& $-0.734\pm0.060$  
& $-1.017\pm0.082$  
& $ $ 
\tabularnewline
\hline 
$\Lambda\rightarrow p$  
& $0.681\pm0.035$  
& $0.896\pm0.043$  
& $ $ 
\tabularnewline
$\Sigma^{-}\rightarrow n$  
& $-0.734\pm0.060$  
& $-1.017\pm0.082$  
& $-0.97\pm0.14$\phantom{a11} 
\tabularnewline
$\Xi^{0}\rightarrow\Sigma^{+}$  
& $1.389\pm0.042$  
& $1.853\pm0.056$  
& $2.0\pm1.2\pm 0.5$ 
\tabularnewline
$\Xi^{-}\rightarrow\Lambda$  
& $-0.026\pm0.042$  
& $-0.060\pm0.056$  
& $ $\tabularnewline
$\Xi^{-}\rightarrow\Sigma^{0}$  
& $1.389\pm0.042$  
& $1.853\pm0.056$  
& $ $ 
\tabularnewline
\hline \hline
\end{tabular}
\caption{Numerical results for the ratios of the vector HSD constants
  $f_{2}/f_{1}$ of the baryon octet. The superscripts ``sym'' and
  ``br'' denote the contribution in exact SU(3) symmetry and the total 
  results with  flavor SU(3) symmetry breaking, respectively. Note
  that the values for the $\Sigma^-\to \Lambda$ mode are given for
  $\sqrt{3/2}f_2$ instead of $f_2/f_1$. }
\label{tab:4}
\end{table}
In Table~\ref{tab:4} we list the present results for the vector HSD 
constants $f_2(0)/f_1(0)$. There are only two existing data for  
the decay processes $\Sigma^-\to n e^- \bar{\nu}_e$ and $\Xi^0 \to
\Sigma^+ e^- \bar{\nu}_e$ from the PDG. The present results are in
agreement with these experimental data, though the experimental
uncertainties of $f_2(0)/f_1(0)$ for the $\Xi^0 \to \Sigma^+ e^-
\bar{\nu}_e$ decay are rather large. Note that 
there are old data for the $\Lambda\to pe^-\bar{\nu}_e$
decay~\cite{Bourquin:1981ba, Bourquin:1983de, Dworkin:1990dd}. In
Ref.~\cite{Dworkin:1990dd}, the vector HSD constant for $\Lambda\to
pe^-\bar{\nu}_e$ decay is given as $f_2(0)/f_1(0)(\Lambda\to
p)=0.15\pm 0.30$. However, this value needs to be scrutinized
experimentally, because firstly its uncertainty is very large and
secondly its value itself is quite different from most theoretical 
predictions. In fact, Cabbibo et al.~\cite{Cabibbo:2003cu} reviewed
the old experimental data~\cite{Wise:1980iq, Wise:1980xx} and
extracted the following values: $f_1=1.238\pm 0.024$ and $f_2=1.34\pm
0.20$, which is in line with the old data: $f_1=1.229\pm
0.035$~\cite{Wise:1980xx}. The prediction from the present work is
$f_2=1.098\pm0.053$, which is closer to that from
Ref.~\cite{Cabibbo:2003cu}. As shown in Table~\ref{tab:4}, the effects
of SU(3) symmetry breaking on the HSD vector constants appear
consistently to be more than $20\,\%$.   

In order to determine the dynamical parameters $a_{i}$ in
eq. (\ref{eq:14}), we use the experimental data for the HSD
constants $g_{1}/f_{1}(B_{1}\rightarrow B_{2})$ listed in
Table~\ref{tab:HSDPDG}. Since there exist only five experimental data for
them as listed in Table~\ref{tab:HSDPDG} but six unknown parameters
$a_i$, we need at least one more data. There are three known
diagonal axial-vector constants, that is, $ g_A^{(0)}$, $g_A^{(3)}$,
and $g_A^{(8)}$. However, the triplet axial-vector constant
$g_A^{(3)}$ is the same as that for the neutron 
$\beta$ decay $g_1/f_1(n\to p)$, we have to avoid it. The
octet one $g_A^{(8)}$ is usually determined by using the HSD constants
with SU(3) symmetry assumed, it is also not suitable for the present
purpose. The rest is the singlet axial-vector constant $g_A^{(0)}$, which
implies the quark content of the nucleon spin and is determined by the
data for polarized electron-proton deep inelastic scattering. Thus, in
addition to the experimental data for the five known HSD constants, we
utilize $g_{A}^{(0)}$ to fix $a_i$. 

\begin{table}[htp]
\begin{tabular}{llc}
\hline 
\hline
    & Experimental data 
& References
\tabularnewline
\hline 
 $g_{1}/f_{1}\left(n\to p\right)$ 
& $1.2701\pm0.0025$ 
& PDG~\cite{Agashe:2014kda}\\
 $g_{1}/f_{1}\left(\Lambda\to p\right)$ 
& $0.718\pm0.015$ 
& PDG~\cite{Agashe:2014kda} \\
 $g_{1}/f_{1}\left(\Sigma^{-}\to n\right)$ 
& $-0.340\pm0.017$ 
& PDG~\cite{Agashe:2014kda}\\
 $g_{1}/f_{1}\left(\Xi^{-}\to\Lambda\right)$ 
& $0.25\pm0.05$ 
& PDG~\cite{Agashe:2014kda} \\
$g_{1}/f_{1}\left(\Xi^{0}\rightarrow\Sigma^{+}\right)$ 
& $1.21\pm0.05$ 
& PDG~\cite{Agashe:2014kda} \\
$g_{A}^{0}$ 
& $0.36\pm0.03$ 
& Bass et al.~\cite{Bass:2004xa} \\
\hline \hline
\end{tabular}
\caption{The experimental data for the HSD constants from the Particle
  Data Group~\cite{Agashe:2014kda}. } 
\label{tab:HSDPDG}
\end{table}
With these six input data from experiments at hand, the
dynamical paramters $a_{i}$ can be determined by solving the 
matrix equation similar to eq.(\ref{eq:MatrixEq16}). The parameters
$a_{i}$ determined by eq.~(\ref{eq:a}) are listed in
Table~\ref{tab:axial_parameter}.  
\begin{table}[htp]
\begin{tabular}{cccccc}
\hline \hline
$a_{1}$ 
& $a_{2}$ 
& $a_{3}$ 
& $a_{4}$ 
& $a_{5}$ 
& $a_{6}$
\\
\hline
$-3.51\pm0.01$ 
& $3.44\pm0.03$ 
& $0.60\pm0.03$ 
& $-1.21\pm0.07$ 
& $0.48\pm0.02$ 
& $-0.74\pm0.04$
\\
\hline \hline
\end{tabular}
\caption{The dynamical parameters of axial-vector transitions}
\label{tab:axial_parameter} 
\end{table}
Having fixed $a_i$, we can straightforwardly solve the
following matrix equation to find other HSD constants and $g_A^8$: 
\begin{eqnarray}
\left[\begin{array}{c}
g_{1}/f_{1}\left(n\rightarrow p\right)\\
g_{1}/f_{1}\left(\Sigma^{+}\rightarrow\Lambda\right)\\
g_{1}/f_{1}\left(\Lambda\rightarrow p\right)\\
g_{1}/f_{1}\left(\Sigma^{-}\rightarrow n\right)\\
g_{1}/f_{1}\left(\Xi^{-}\rightarrow\Lambda\right)\\
g_{1}/f_{1}\left(\Xi^{-}\rightarrow\Sigma^{0}\right)\\
g_{1}/f_{1}\left(\Sigma^{-}\rightarrow\Lambda\right)\\
g_{1}/f_{1}\left(\Sigma^{-}\rightarrow\Sigma^{0}\right)\\
g_{1}/f_{1}\left(\Xi^{-}\rightarrow\Xi^{0}\right)\\
g_{1}/f_{1}\left(\Xi^{0}\rightarrow\Sigma^{+}\right)\\
g_{A}^{0}(p)\\
g_{A}^{3}(p)\\
g_{A}^{8}(p)
\end{array}\right] 
& = & 
\mathbf{M}_A\;\left[\begin{array}{c}
a_{1}\\
a_{2}\\
a_{3}\\
a_{4}\\
a_{5}\\
a_{6}
\end{array}\right],
\label{eq:a}
\end{eqnarray}
where the matrix $\mathbf{M}_A$ is expressed as
 \begin{equation}
{ \displaystyle \hspace{-4em}\mathbf{M}_A
\;=\;
\left[\begin{array}{cccccc}
-\frac{7}{30} - \frac{1}{3}c_{\overline{10}} - \frac{2}{45}c_{27} 
& \frac{7}{60} - \frac{1}{3}c_{\overline{10}} - \frac{4}{45}c_{27} 
& \frac{1}{60} - \frac{1}{6}c_{\overline{10}} + \frac{1}{15}c_{27} 
& -\frac{11}{270} & -\frac{1}{9} 
& -\frac{1}{15}\\[0.3em]
-\frac{3}{20} - \frac{1}{4}c_{\overline{10}} - \frac{1}{12}c_{27} 
& \frac{3}{40} - \frac{1}{4}c_{\overline{10}} - \frac{1}{6}c_{27} 
& -\frac{1}{40} - \frac{1}{8}c_{\overline{10}} + \frac{1}{8}c_{27} 
& -\frac{7}{180} 
& -\frac{1}{30} 
& 0\\[0.3em]
-\frac{2}{15} + \frac{1}{6}c_{\overline{10}} + \frac{1}{30}c_{27} 
& \frac{1}{15} + \frac{1}{6}c_{\overline{10}} + \frac{1}{15}c_{27} 
& \frac{1}{30} + \frac{1}{12}c_{\overline{10}} - \frac{1}{20}c_{27} 
& \frac{1}{45} 
& 0 
& -\frac{1}{30}\\[0.3em]
\frac{1}{15} - \frac{1}{45}c_{27} 
& -\frac{1}{30} - \frac{2}{45}c_{27} 
& \frac{1}{15} + \frac{1}{30}c_{27} 
& -\frac{1}{270} 
& -\frac{1}{45} 
& \frac{1}{30}\\[0.3em]
-\frac{1}{30} - \frac{1}{30}c_{27} 
& \frac{1}{60} - \frac{1}{15}c_{27} 
& \frac{1}{20} + \frac{1}{20}c_{27} 
& -\frac{1}{180} 
& \frac{1}{30} 
& -\frac{1}{30}\\[0.3em]
-\frac{7}{30} + \frac{1}{6}c_{\overline{10}}
+\frac{1}{45}c_{27} 
& \frac{7}{60} + \frac{1}{6}c_{\overline{10}}
+\frac{2}{45}c_{27} 
& \frac{1}{60} + \frac{1}{12}c_{\overline{10}}
-\frac{1}{30}c_{27} 
& \frac{11}{540} 
& \frac{1}{18} 
& \frac{1}{30}\\[0.3em]
-\frac{3}{20} - \frac{1}{4}c_{\overline{10}}
-\frac{1}{12}c_{27} 
& \frac{3}{40} - \frac{1}{4}c_{\overline{10}}
-\frac{1}{6}c_{27} 
& -\frac{1}{40} - \frac{1}{8}c_{\overline{10}}
+\frac{1}{8}c_{27} 
& -\frac{7}{180} 
& -\frac{1}{30} 
& 0
\\[0.3em]
-\frac{1}{12}
-\frac{1}{6}c_{\overline{10}} 
&
\frac{1}{24}
-\frac{1}{6}c_{\overline{10}} 
& 
\frac{1}{24}
-\frac{1}{12}c_{\overline{10}} 
& 
-\frac{1}{60} 
& 
-\frac{1}{30} 
& 
\frac{1}{30}\\[0.3em]
\frac{1}{15}
+\frac{2}{45}c_{27} 
& 
-\frac{1}{30}
+\frac{4}{45}c_{27} 
& 
\frac{1}{15}
-\frac{1}{15}c_{27} 
& 
\frac{1}{135} 
& 
\frac{2}{45} 
& 
-\frac{1}{15}
\\[0.3em]
-\frac{7}{30}
+\frac{1}{6}c_{\overline{10}}
+\frac{1}{45}c_{27} 
& 
\frac{7}{60}
+\frac{1}{6}c_{\overline{10}}
+\frac{2}{45}c_{27} 
& 
\frac{1}{60}
+\frac{1}{12}c_{\overline{10}}
-\frac{1}{30}c_{27} 
& 
\frac{11}{540} 
& \frac{1}{18} 
& \frac{1}{30}
\\[0.3em]
0 
& 
0 
& 
1 
& 
0 
& 
-\frac{1}{5} 
& 
\frac{1}{5}
\\[0.3em]
-\frac{7}{30}
-\frac{1}{3}c_{\overline{10}}
-\frac{2}{45}c_{27} 
& 
\frac{7}{60}
-\frac{1}{3}c_{\overline{10}}
-\frac{4}{45}c_{27} 
& 
\frac{1}{60}
-\frac{1}{6}c_{\overline{10}}
+\frac{1}{15}c_{27} 
& 
-\frac{11}{270} 
& 
-\frac{1}{9} 
& 
-\frac{1}{15}\cr
-\frac{1}{10\sqrt{3}}
+\frac{1}{\sqrt{3}}c_{\overline{10}}
-\frac{2}{5\sqrt{3}}c_{27}   
&
\frac{1}{20\sqrt{3}}
+\frac{1}{\sqrt{3}}c_{\overline{10}}
-\frac{4}{5\sqrt{3}}c_{27}   
&
\frac{\sqrt{3}}{20}
+\frac{1}{2\sqrt{3}}c_{\overline{10}}
+\frac{\sqrt{3}}{5}c_{27}  
& \frac{1}{30\sqrt{3}} 
& 
0 
& 
0
\end{array}\right].}
\label{eq:matrixMA}
\end{equation}

In the exact $\mathrm{SU(3)}$ symmetry, all axial-vector decay
amplitudes are given in terms of two reduced matrix elements $F$ and
$D$ as follows: 
\begin{eqnarray}
g_{1}^{(B_1\to B_2)}(0) 
& = & 
C_{F}^{(B_1\to  B_2)}F
\;+\; 
C_{D}^{(B_1\to B_2)}D.
\label{eq:g1_CF_CD}
\end{eqnarray}
Here $C_{F}^{(B\rightarrow B^{\prime})}$ and $C_{D}^{(B\rightarrow
B^{\prime})}$ are SU(3) Clebsch-Gordan coefficients that appear when
the axial-vector collective operator $\hat{g}_1^{B_1\to B_2}$ in
eq.(\ref{eq:14}) is sandwiched between octet states. Thus, the $F$ and
$D$ are determined in terms of $a_1$, $a_2$, and $a_3$
\begin{eqnarray}
F\;\;=\;\;-\frac{1}{12}\left(a_{1}-\frac{1}{2}a_{2}\right)
  +\frac{1}{24}a_{3}   
& = & 0.461\pm0.002,\cr 
D\;\;=\;\;-\frac{3}{20}\left(a_{1}-\frac{1}{2}a_{2}\right)
      -\frac{1}{40}a_{3}  
& = & 0.769\pm0.003.
\label{eq:FD}
\end{eqnarray}
The results for the $F$ and $D$ are in good agreement with the
empirical ones extracted from the experiments~\cite{Goto:1999by},
given as $F=0.463\pm 0.008$ and $D=0.804\pm0.008$.  

\begin{table}[htp]
\begin{tabular}{lccc}
\hline \hline
Decay mode  
& $f_{1}(0)$  
& $g_{1}(0)$  
& $g_{1}(0)/f_{1}(0)$ 
\\[0.3em]
\hline 
$n\rightarrow p$  
& $1$  
& $F+D$  
& $F+D$ 
\\[0.3em]
$\Sigma^{-}\rightarrow\Lambda$  
& $0$
& $\sqrt{\frac{2}{3}}D$  
& $D$
\\[0.3em]
$\Sigma^{0}\rightarrow\Sigma^{+}$  
& $\sqrt{2}$
& $\sqrt{2}F$  
& $F$ 
\\[0.3em]
$\Sigma^{-}\rightarrow\Sigma^{0}$  
& $\sqrt{2}$
& $\sqrt{2}F$  
& $F$ 
\\[0.3em]
$\Xi^{-}\rightarrow\Xi^{0}$ 
& $1$
& $F-D$  
& $F-D$ 
\\[0.3em]
\hline 
$\Lambda\rightarrow p$  
& $\sqrt{\frac32}$
& $\sqrt{\frac{3}{2}}\left(F+\frac{1}{3}D\right)$  
& $F+\frac{1}{3}D$ 
\\[0.3em]
$\Sigma^{0}\rightarrow p$  
& $\frac1{\sqrt{2}}$
& $\frac{1}{\sqrt{2}}\left(F-D\right)$  
& $F-D$ 
\\[0.3em]
$\Sigma^{-}\rightarrow n$  
& $1$
& $F-D$  
& $F-D$ 
\\[0.3em]
$\Xi^{0}\rightarrow\Sigma^{+}$  
& $1$
& $F+D$  
& $F+D$ 
\\[0.3em]
$\Xi^{-}\rightarrow\Lambda$  
& $\sqrt{\frac32}$
& $ \sqrt{\frac{3}{2}}\left(F-\frac{1}{3}D\right)$  
& $F-\frac{1}{3}D$ 
\\[0.3em]
$\Xi^{-}\rightarrow\Sigma^{0}$  
& $\frac1{\sqrt{2}}$
& $\frac{1}{\sqrt{2}}\left(F+D\right)$  
& $F+D$ 
\\[0.3em]
\hline \hline
\end{tabular}
\caption{The expressions of $g_{1}(0)$ and $g_1(0)/f_1(0)$ in exact
  SU(3) symmetry. }  
\label{tab:SU3Axial}
\end{table}

\begin{table}[ht]
\begin{tabular}{lcccc}
\hline 
Decay modes
& 
$(g_1/f_1)^{\mathrm{sym}}$ 
& $(g_1/f_1)^{\mathrm{br}}$
& Exp. (Input) 
& Refs.
\tabularnewline  
\hline 
$n\to p$ 
& $1.230\pm0.004$ 
& $1.269\pm0.006$ 
& $1.2701\pm0.0025$ 
& \cite{Agashe:2014kda}
\tabularnewline 
 $\Sigma^{-}\to\Lambda$ 
& $0.769\pm0.003$ 
& $0.794\pm0.004$ 
&  
& 
\tabularnewline
 $\Sigma^{-}\to\Sigma^{0}$ 
& $0.461\pm0.002$ 
& $0.439\pm0.003$ 
&  
& 
\tabularnewline
 $\Xi^{-}\to\Xi^{0}$ 
& $-0.308\pm0.002$ 
& $-0.245\pm0.004$ 
&  
& 
\tabularnewline
 $\Lambda\to p$ 
& $0.717\pm0.003$ 
& $0.718\pm0.003$ 
& $0.718\pm0.015$ 
& \cite{Agashe:2014kda}
\tabularnewline
 $\Sigma^{-}\to n$ 
& $-0.308\pm0.002$ 
& $-0.340\pm0.003$
& $-0.340\pm0.017$ 
& \cite{Agashe:2014kda}
\tabularnewline
 $\Xi^{0}\rightarrow\Sigma^{+}$ 
& $1.230\pm0.004$ 
& $1.210\pm0.005$ 
& $1.21\pm0.05$ 
& \cite{Agashe:2014kda}
\tabularnewline
 $\Xi^{-}\to\Lambda$ 
& $0.204\pm0.002$ 
& $0.250\pm0.002$ 
& $0.25\pm0.05$ 
& \cite{Agashe:2014kda}
\tabularnewline
 $\Xi^{-}\to\Sigma^{0}$ 
& $1.230\pm0.004$ 
& $1.210\pm0.005$ 
& & 
\tabularnewline
$g_{A}^{(0)}$ 
& $0.604\pm0.030$ 
& $0.361\pm0.031$ 
& $0.36\pm0.03$
&
\cite{Bass:2004xa} 
\tabularnewline
 $g_{A}^{(8)}$ 
& $0.354\pm0.003$ 
& $0.325\pm0.004$ 
&  
& 
\tabularnewline
\hline 
\end{tabular}
\caption{Numerical results for the ratios of the axial-vector HSD 
  constants $g_1/f_{1}$ of the baryon octet. The superscripts ``sym''
  and ``br'' denote the contribution in exact SU(3) symmetry and the
  total results with  flavor SU(3) symmetry breaking,
  respectively. Note that the values for the $\Sigma^-\to \Lambda$
  mode are given for $\sqrt{3/2}g_1$ instead of $g_1/f_1$. The
  experimental data are taken from the Particle Data
  Group~\cite{Agashe:2014kda}, which are used for the input.}
\label{tab:axialHSD}
\end{table}
In Table~\ref{tab:axialHSD}, the results for the axial-vector HSD
constants $g_1/f_1$ are listed. In contrast to the results for
$f_2/f_1$, the effects of flavor SU(3) symmetry breaking on the
axial-vector HSD constants are rather small and stable. However, when
it comes to the singlet axial charge, they are not negligible. Note
that though $g_A^{(0)}$ is used as an input, we still can estimate the
effects of SU(3) symmetry breaking, which turn out to be approximately
$40\,\%$. The reason why the contribution of SU(3) symmetry breaking
is large in the case the singlet axial charge can be found in
eqs.(\ref{eq:a}, \ref{eq:matrixMA}). The expression for $g_A^{(0)}$ is
written as 
\begin{equation}
  \label{eq:ga0}
g_A^{(0)}\;=\; a_3 -\frac15(a_5-a_6).  
\end{equation}
The singlet axial charge does not have any contribution from the
leading order in the $1/N_c$ expansion. Indeed, this explains
why the Skyrme model predicts the smallness of $g_A^{(0)}$. Because of
this fact, there is no wavefunction corrections to
$g_A^{(0)}$. Moreover, the relative signs of $a_5$ and $a_6$ are
different, so that the linear $m_s$ corrections turn out to be rather
large. Thus, the effects of SU(3) symmetry breaking play an essential
role in describing the quark content of the nucleon spin. 

The octet axial-vector constant $g_A^{(8)}$ was usually extracted from
the HSD data with flavor SU(3) symmetry assumed and its value was
obtained to be $g_A^{(8)}=0.58\pm0.03$~\cite{Close:1993mv}. On the
other hand, the present result for $g_A^{(8)}$ in the SU(3) symmetric
case is given as $g_A^{(8)}=0.354\pm 0.003$. With the linear $m_s$
corrections taken into account, the value of $g_A^{(8)}$ is reduced to  
$g_A^{(8)}=0.325\pm0.004$, which is not much different from that with
SU(3) symmetry. However, compared with the extracted value from
Ref.~\cite{Close:1993mv}, the present result for $g_A^{(8)}$ is
approximately $40\,\%$ smaller than that. Note that the octet
axial charge is expressed in terms of $F$ and $D$ as follows: 
\begin{equation}
  \label{eq:OctetGa}
g_A^{(8)} \;=\; \frac1{\sqrt{3}} (3F-D) =
\frac1{\sqrt{3}}\left[-\frac1{10}(a_1-\frac12 a_2) +
  \frac1{20}a_3\right]. 
\end{equation}
Inserting the values of $F$ and $D$ given in Ref.~\cite{Goto:1999by},
we find $g_A^{(8)}=0.338\pm0.015$, which is quite different from that
of Ref.~\cite{Close:1993mv} but is compatible with the present value.  
\section{Analysis for the transition from the baryon decuplet to the
  octet}
In this Section, we want to present the results for the axial-vector
transition constants of the baryon decuplet. Since the baryon decuplet
except for $\Omega^-$ decays into the baryon octet strongly, there are
no direct experimental data for the axial-vector HSD constants for the
baryon decuplet to date. However, there are at least two evident
reasons why the axial-vector transition constants for the baryon
decuplet are interesting. Firstly, the baryon decuplet can be produced
by exclusive neutrino-nucleon scattering such as $\nu+ N\to \mu+
\Delta$ and $\nu+ N\to \mu+ B_{10}+M_8$ processes, where $M_8$ stands
for the pseudo-scalar meson octet. These reactions are also
interesting, since they may bring out information on weak 
generalized parton distributions~\cite{Psaker:2006gj}. The
axial-vector transition constants for the baryon decuplet might be
extracted from these reactions. Secondly, using the Goldberger-Treiman
relation, one can relate the axial-vector transition constants to the 
strong coupling constants for vertices of the baryon
decuplet-octet and the meson octet. Because of these reasons, it is of
great interest to understand the axial-vector transition constants for
the baryon decuplet.

There are four different axial-vector form factors or Adler form
factors $C_i^A(q^2)$ ($i=3,\,4,\,5,\,6$) as given in
eq.(\ref{eq:Adler}). However, $C_3^A$ and $C_4^A$ are often
neglected, since the partially conserved axial-vector current (PCAC)
relation yields an expression, where the coefficients in front of
$C_3^A$ and $C_4^A$ are proportional to the mass difference of a
decuplet baryon and a octet one divided by the mass sum, and its
squares, respectively~\cite{Berman:1965iu}. Thus, their contribution
is negligible in the transition amplitude of $\nu N\to \mu B_{10}$
scattering. In the chiral limit, $C_6^A$ can be related to $C_5^A$, so
that we will concentrate on $C_5^A$ in this work.   

Before we present the results, we want to discuss possible relations
and sum rules for the axial-vector transition constants $C_5^A$. With
the linear 
$m_{\mathrm{s}}$ corrections turned on, we find the following 
eight relations for $C_5^A$: 
\begin{eqnarray}
 \left(\Delta^{0}\ensuremath{\to}p\right) 
& = & 
 \frac{1}{\sqrt{3}}\left(\Delta^{-}\rightarrow n\right)
 \;=\;
- \frac{1}{\sqrt{3}}\left(\Delta^{\text{++}}\rightarrow p\right)
\;=\;
-\left(\Delta^{+}\rightarrow n\right),\nonumber\\[0.3em]
 \left(\Sigma^{\ast0}\rightarrow\Sigma^{+}\right) 
& = & 
 \left(\Sigma^{\ast-}\rightarrow\Sigma^{0}\right)
\;=\;
 \left(\Sigma^{\ast+}\rightarrow\Sigma^{0}\right)
\;=\;
 \left(\Sigma^{\ast0}\rightarrow\Sigma^{-}\right),
\nonumber\\[0.3em]
\left(\Sigma^{\ast-}\rightarrow\Lambda\right)
& = & 
\left(\Sigma^{\ast+}\rightarrow\Lambda\right),
\nonumber\\[0.3em]
  \left(\Xi^{\ast-}\rightarrow\Xi^{0}\right)
& = & 
\left(\Xi^{\ast0}\to\Xi^{-}\right),
\nonumber\\[0.3em]
  \left(\Delta^{++}\rightarrow\Sigma^{+}\right) 
& = & 
\sqrt{\frac{3}{2}}\left(\Delta^{+}\rightarrow\Sigma^{0}\right)
\;=\;\sqrt{3}\left(\Delta^{0}\rightarrow\Sigma^{-}\right),
\nonumber\\[0.3em]
\left(\Sigma^{\ast0}\rightarrow p\right)
& = & 
\frac{1}{\sqrt{2}}\left(\Sigma^{\ast-}\rightarrow n\right),
\nonumber\\[0.3em]
 \left(\Sigma^{\ast+}\rightarrow\Xi^{0}\right)
 &=&
 \sqrt{2}\left(\Sigma^{\ast0}\rightarrow\Xi^{-}\right),
\nonumber\\[0.3em]
\left(\Xi^{\ast0}\rightarrow\Sigma^{+}\right)
& = & 
 \sqrt{2}\left(\Xi^{\ast-}\rightarrow\Sigma^{0}\right).
\label{eq:relation}
\end{eqnarray}
These relations come from isospin symmetry. We also obtain the
following six sum rules for the $C_5^A$:
\begin{eqnarray}
 \left(\Delta^{0}\ensuremath{\to}p\right) 
 & = & 
- \left(\Xi^{\ast-}\to\Xi^{0}\right)
- \frac{1}{\sqrt{2}}\left(\Xi^{*-}\rightarrow\Sigma^{0}\right)
+ \sqrt{\frac{3}{2}}\left(\Xi^{\ast-}\rightarrow\Lambda\right)
+ \frac{2}{\sqrt{3}}\left(\Omega^{-}\rightarrow\Xi^{0}\right),
\nonumber\\[0.3em]
 \left(\Sigma^{\ast0}\rightarrow\Sigma^{+}\right) 
 & = & 
 \frac{1}{\sqrt{2}}\left(\Xi^{\ast-}\rightarrow\Xi^{0}\right)
+ \left(\Xi^{\ast-}\rightarrow\Sigma^{0}\right)
- \frac{1}{\sqrt{6}}\left(\Omega^{-}\rightarrow\Xi^{0}\right),
\nonumber\\[0.3em]
\left(\Sigma^{\ast-}\rightarrow\Lambda\right)
& = & 
- \sqrt{\frac{3}{2}}\left(\Xi^{\ast-}\rightarrow\Xi^{0}\right)
+ \left(\Xi^{\ast-}\rightarrow\Lambda\right)
+ \frac{1}{\sqrt{2}}\left(\Omega^{-}\rightarrow\Xi^{0}\right),
\nonumber\\[0.3em]
  \left(\Xi^{\ast-}\rightarrow\Xi^{0}\right)
& = & 
  \sqrt{\frac{2}{3}}\left(\Xi^{\ast-}\rightarrow\Lambda\right)
- \sqrt{\frac{2}{3}}\left(\Sigma^{\ast-}\rightarrow\Lambda\right)
+ \frac{1}{\sqrt{3}}\left(\Omega^{-}\rightarrow\Xi^{0}\right),
\nonumber\\[0.3em]
\left(\Sigma^{\ast0}\rightarrow p\right)
& = & 
- \frac{1}{2}\left(\Xi^{\ast-}\rightarrow\Sigma^{0}\right)
+ \frac{\sqrt{3}}{2}\left(\Xi^{\ast-}\rightarrow\Lambda\right)
+ \frac{1}{\sqrt{6}}\left(\Omega^{-}\rightarrow\Xi^{0}\right),
\nonumber\\[0.3em]
  \left(\Delta^{++}\rightarrow\Sigma^{+}\right) 
& = & 
- \sqrt{6}\left(\Xi^{\ast-}\rightarrow\Sigma^{0}\right)
+ \sqrt{6}\left(\Sigma^{\ast0}\rightarrow\Xi^{-}\right)
+ \left(\Omega^{-}\rightarrow\Xi^{0}\right).
\label{eq:sumrule1}
\end{eqnarray}
Though we do not expect that the sum rules in eq.(\ref{eq:sumrule1})
will be confirmed by the experimental data in near future, they
provide a consistency check of the present results. 

\begin{table}[htp]
\begin{tabular}{cccc}
\hline 
\noalign{\vskip\doublerulesep}
$B_{10}\overset{X=1+i2}{\rightarrow}B_{8}$ 
& 
${\displaystyle
    C_{5}^{A\,(\mathrm{sym})}}$
&   $C_{5}^{A\,(\mathrm{br})}$
\tabularnewline[\doublerulesep]
\hline 
$
\begin{array}{c}
\Delta^{0}\to p\\
\Delta^{-}\to n\\
\Sigma^{\ast0}\to\Sigma^{+}\\
\Sigma^{\ast-}\to\Sigma^{0}\\
\Sigma^{\ast-}\to\Lambda\\
\Xi^{\ast-}\to\Xi^{0}
\end{array}
$ 
& 
$
\begin{array}{r}
-0.954\pm0.003\\
-1.653\pm0.006\\
0.675\pm0.002\\
0.675\pm0.002\\
-1.169\pm0.004\\
0.954\pm0.003
\end{array}
$ 
& 
$\begin{array}{r}
-1.040\pm0.005\\
-1.801\pm0.008\\
0.614\pm0.004\\
0.614\pm0.004\\
-1.231\pm0.005\\
0.903\pm0.006
\end{array}$
\tabularnewline
\hline 
\end{tabular}
\caption{Axial-vector transition constants of the baryon decuplet to
 the octet with $X=1+i2$, i.e. in the strangeness-conserving ($\Delta
 S=0$) case. The charge difference between the initial state and the
 final state is $\Delta Q=+1$. }
\label{tab:NumImDecOct1}
\end{table}
\begin{table}
\begin{tabular}{cccc}
\hline 
\noalign{\vskip\doublerulesep}
    $B_{10}\overset{X=1-i2}{\rightarrow}B_{8}$ 
& 
    ${\displaystyle C_{5}^{A\,(\mathrm{sym})}}$
&   $C_{5}^{A\,(\mathrm{br})}$
\tabularnewline[\doublerulesep]
\hline 
$\begin{array}{c}
\Delta^{++}\to p\\
\Delta^{+}\to n\\
\Sigma^{\ast+}\to\Sigma^{0}\\
\Sigma^{\ast+}\to\Lambda\\
\Sigma^{\ast0}\to\Sigma^{-}\\
\Xi^{\ast0}\to\Xi^{-}
\end{array}$ 
&  
$\begin{array}{r}
1.653\pm0.006\\
0.954\pm0.003\\
0.675\pm0.002\\
1.169\pm0.004\\
0.675\pm0.002\\
0.954\pm0.003
\end{array}$ 
&  
$\begin{array}{r}
1.801\pm0.008\\
1.040\pm0.005\\
0.614\pm0.004\\
1.231\pm0.005\\
0.614\pm0.004\\
0.903\pm0.006
\end{array}$
\tabularnewline
\hline 
\end{tabular}
\caption{Axial-vector transition constants of the baryon decuplet to
  octet with the flavor transition operator of $X\,=\,1 - i\,2$,
  i.e. in the strangeness-conserving ($\Delta S=0$) case. The charge
  difference between the initial state and the  final state is $\Delta
  Q=-1$.} 
\label{tab:NumImDecOct2}
\end{table}

In Tables~\ref{tab:NumImDecOct1} and \ref{tab:NumImDecOct2}, we list
the results for the axial-vector transition constants of the baryon
decuplet decaying into the octet in the strangeness-conserving
cases ($\Delta S=0$). Table~\ref{tab:NumImDecOct1} presents those for
the positive charge difference between the initial state and the
final state ($\Delta Q=+1$), and Table~\ref{tab:NumImDecOct2} for the
negative charge difference ($\Delta Q=-1$). 
Tables~\ref{tab:NumVmDecOct1} and \ref{tab:NumVmDecOct2} list the
results for $C_5^A$ in the strangeness-changing cases ($\Delta S=1$).
Note that all the results presented in Tables~\ref{tab:NumImDecOct1},
\ref{tab:NumImDecOct2}, \ref{tab:NumVmDecOct1} and
\ref{tab:NumVmDecOct2} satisfy the isospin relations given in
eq.(\ref{eq:relation}) and also the sum rules in
eq.(\ref{eq:sumrule1}). The effects of flavor SU(3) symmetry breaking
turn out to be rather small in the case of $C_5^A$. 
\begin{table}[htp]
\begin{tabular}{cccc}
\hline 
\noalign{\vskip\doublerulesep}
$B_{10}\overset{X=4+i5}{\rightarrow}B_{8}$
& 
${\displaystyle
    C_{5}^{A\,(\mathrm{sym})}}$
&   $C_{5}^{A\,(\mathrm{br})}$
\tabularnewline[\doublerulesep]
\hline 
$\begin{array}{c}
\Sigma^{\ast0}\to p\\
\Sigma^{\ast-}\to n\\
\Xi^{\ast0}\to\Sigma^{+}\\
\Xi^{\ast-}\to\Sigma^{0}\\
\Xi^{\ast-}\to\Lambda\\
\Omega^{-}\to\Xi^{0}
\end{array}$ 
&  $\begin{array}{r}
-0.675\pm0.002\\
-0.954\pm0.003\\
0.954\pm0.003\\
0.675\pm0.002\\
-1.169\pm0.004\\
1.653\pm0.006
\end{array}$ 
& $\begin{array}{r}
-0.755\pm0.004\\
-1.067\pm0.005\\
0.896\pm0.004\\
0.633\pm0.003\\
-1.266\pm0.005\\
1.612\pm0.007
\end{array}$
\tabularnewline
\hline 
\end{tabular}
\caption{Axial-vector transition constants of the baryon decuplet to
 the octet with $X=4+i5$, i.e. in the strangeness-changing ($\Delta
 S=1$) case. The charge difference between the initial state and the
 final state is $\Delta Q=+1$.}
\label{tab:NumVmDecOct1}
\end{table}

\begin{table}[htp]
\begin{tabular}{cccc}
\hline 
\noalign{\vskip\doublerulesep}
$B_{10}\overset{X=4-i5}{\rightarrow}B_{8}$
&
${\displaystyle
    C_{5}^{A\,(\mathrm{sym})}}$
&  $C_{5}^{A\,(\mathrm{br})}$
\tabularnewline[\doublerulesep]
\hline 
 $\begin{array}{c}
\Delta^{++}\to\Sigma^{+}\\
\Delta^{+}\to\Sigma^{0}\\
\Delta^{+}\to\Lambda\\
\Delta^{0}\to\Sigma^{-}\\
\Sigma^{\ast+}\to\Xi^{0}\\
\Sigma^{\ast0}\to\Xi^{-}
\end{array}$ 
&  $\begin{array}{r}
-1.653\pm0.006\\
-1.350\pm0.005\\
0\\
-0.954\pm0.003\\
-0.954\pm0.003\\
-0.675\pm0.002
\end{array}$ 
&  $\begin{array}{r}
-1.547\pm0.007\\
-1.263\pm0.005\\
0\\
-0.893\pm0.004\\
-0.928\pm0.004\\
-0.656\pm0.003
\end{array}$
\tabularnewline
\hline 
\end{tabular}
\caption{Axial-vector transition constants of the baryon decuplet to
 the octet with $X=4-i5$, i.e. in the strangeness-changing ($\Delta
 S=1$) case. The charge difference between the initial state and the
 final state is $\Delta Q=-1$.}
\label{tab:NumVmDecOct2}
\end{table}

In the SU(6) quark model, the $C_5^A(n\to \Delta^+)$ can be related to
the axial-vector HSD constant $g_1/f_1(n\to
p)$~\cite{Albright:1965zz,Llewellyn Smith:1971zm} as 
\begin{equation}
  \label{eq:SU6}
C_5^A(\Delta^+\to n) \;=\; \frac{2\sqrt{3}}{5} g_1/f_1(n\to p).  
\end{equation}
Equation~(\ref{eq:SU6}) has been often used to determine the $\pi
N\Delta$ coupling constant. Thus, it is interesting to examine it
based on the present result. Using the experimental data for
$g_1/f_1(n\to p)$, we obtain $C_5^A(\Delta^+\to n) = 0.88$, whereas
the present result from Table~\ref{tab:NumImDecOct2} is
$C_5^A(\Delta^+\to n) = 0.95$. Thus, the value from the SU(6) relation
is deviated from the present one approximately by $10\,\%$. 

\section{Summary and outlook}
In the present work, we aimed at investigating the hyperon
semileptonic decay constants of the baryon octet within the framework
of a chiral soliton model, determining all the dynamical parameters by
the experimental data. We first reviewed the analysis of the mass
splittings of the baryon octet and decuplet in brief. Then, we 
studied the vector hyperon semileptonic decay constants $f_2/f_1$ with
the dynamical parameters $w_i$ fixed in an unambiguous manner by the
experimental data for the magnetic moments of the baryon octet. we
found that the effects of flavor SU(3) symmetry breaking contributed
to the $f_2/f_1$ approximately by $25\,\%$.

The dynamical parameters $a_i$ were fixed by using the experimental
data for the axial-vector hyperon semileptonic decay constants as well
as the singlet axial charge $g_A^{(0)}$ unequivocally. We first
derived the $F$ and $D$ in exact SU(3) symmetry and the results were
consistent with the empirical data. We also predicted other $g_1/f_1$
which are not measured yet. The octet $g_A^{(8)}$ was obtained to be  
$g_A^{(8)}=0.325\pm 0.004$. The contribution of flavor SU(3) symmetry
breaking turned out to be rather small in the case of the axial-vector
hyperon semileptonic decay constants. 
Finally, we predicted the axial-vector constants for the
transitions from the baryon decuplet to the octet. In addition, we
derived the isospin relations and their six different sum rules. 

The present results for the axial-vector transition constants for the
baryon decuplet can be used to determine the meson-baryon Yukawa
coupling constants by the Goldberger-Treiman relations. The whole
analysis including the baryon decuplet-decuplet-meson coupling
constants is under way. Last but not least, the vector transition
constants for the baryon decuplet are also interesting and
important. However, experimental information on the $E_2/M_1$
ratio for the baryon decuplet is essential but only that of the
$\Delta$ isobar is known. On the other hand, if one assumes SU(3)
symmetry, it might be possible to study the vector transition
constants for the baryon decuplet. The corresponding investigation
will appear elsewhere.  

\acknowledgments
The present work was supported by Basic Science Research Program 
through the National Research Foundation of Korea funded by the
Ministry of Education, Science and Technology (Grant Numbers:
NRF-2013R1A1A2063590 and NRF-2012R1A1A2001083).

\end{document}